\documentstyle[12pt,epsf,epsfig,wrapfig]{article}
\begin{document}

\begin{center}
{\bfseries}QCD VACUUM TOPOLOGICAL SUSCEPTIBILITY IN A NON-LOCAL CHIRAL QUARK MODEL

\vskip 5mm

\underline{A.E. Dorokhov}\footnote{{\it E-mail: dorokhov@thsun1.jinr.ru }}

\vskip 5mm

Bogoliubov Laboratory of Theoretical Physics, Joint Institute for Nuclear

Research, 141980, Dubna, Russia
\end{center}

\vskip 5mm
\begin{abstract}
The topological susceptibility of QCD vacuum is studied in the framework of a covariant
chiral quark model with non-local quark-quark interaction. The relation of
the first moment of topological susceptibility $\chi ^{\prime }(0)$ and the
'spin crisis' problem is briefly discussed. It is shown, in particular, that
one always gets the inequality $\chi ^{\prime }(0)>\chi _{OZI}^{\prime }$.
\end{abstract}

\vskip 8mm

It is well known that due to $U_{A}\left( 1\right) $ axial Adler-Bell-Jackiw
anomaly the iso-singlet axial-vector current%
\begin{equation}
J_{\mu 5}^{(0)}=\sum_{f}\overline{q}_{f}\gamma _{\mu }\gamma _{5}q_{f}
\label{Jsingl}
\end{equation}%
is not conserved even in the chiral limit, and its divergence equals
\begin{equation}
\partial _{\mu }J_{\mu 5}^{(0)}\left( x\right) =2N_{f}Q_{5}\left( x\right) ,
\label{DivJ0}
\end{equation}%
where%
\begin{equation}
Q_{5}\left( x\right) =(\alpha _{s}/8\pi )G_{\mu \nu }^{a}(x)\tilde{G}_{\mu
\nu }^{a}(x)  \label{Q5x}
\end{equation}%
is the topological charge density. The correlator of singlet currents is
defined as
\begin{eqnarray}
\Pi _{A,\mu \nu }^{(0)}(q) &=&i\int d^{4}x~e^{iqx}\langle 0\left\vert
T\left\{ J_{\mu 5}^{(0)}(x)J_{\nu 5}^{(0)}(0)^{\dagger }\right\} \right\vert
0\rangle =  \nonumber \\
&=&\,\left( q_{\mu }q_{\nu }-g_{\mu \nu }q^{2}\right) \Pi
_{A,T}^{(0)}(Q^{2})+q_{\mu }q_{\nu }\Pi _{A,L}^{(0)}(Q^{2}).  \label{PA}
\end{eqnarray}%
In the chiral limit the longitudinal part of the correlator defines the
topological susceptibility, {\it i.e.} the correlator of the topological
charge densities, $Q_{5}\left( x\right) $,
\begin{equation}
\chi \left( Q^{2}\right) =i\int d^{4}x~e^{iqx}\langle 0\left\vert T\left\{
Q_{5}(x)Q_{5}(0)\right\} \right\vert 0\rangle ,  \label{ChiQ2}
\end{equation}%
with the relation (see, {\em e.g.}, \cite{IofSams00})
\begin{equation}
\Pi _{L}^{A,0}\left( Q^{2}\right) =\frac{\left( 2N_{f}\right) ^{2}}{Q^{2}}%
\chi \left( Q^{2}\right) .  \label{PLchi}
\end{equation}%
At high $Q^{2}$ the operator product expansion (OPE) predicts \cite{SVZeta}%
\begin{equation}
\chi \left( Q^{2}\rightarrow \infty \right) =-\frac{\alpha _{s}}{16\pi }%
\left\langle \frac{\alpha _{s}}{\pi }\left( G_{\mu \nu }^{a}\right)
^{2}\right\rangle +{\cal O}(Q^{-2}),  \label{ChiQCDlarge}
\end{equation}%
where the perturbative contribution has been subtracted.

At low $Q^{2}$, $\chi \left( Q^{2}\right) $ is represented as a sum of
contributions coming purely from QCD and from $\left( \pi ,\eta \right) -$%
mesonic resonances \cite{IofSams00}
\begin{eqnarray}
\left[ \chi \left( Q^{2}\right) \right] _{{\rm full\ QCD}} &=&\frac{%
m_{u}m_{d}}{m_{u}+m_{d}}\left\langle \overline{u}u\right\rangle -\chi
^{\prime }(0)Q^{2}-  \label{ChiQCDsmall} \\
&&-\frac{f_{\pi }^{2}}{4}Q^{2}\left[ \left( \frac{m_{u}-m_{d}}{m_{u}+m_{d}}%
\right) ^{2}\frac{m_{\pi }^{2}}{Q^{2}+m_{\pi }^{2}}+\frac{1}{3}\frac{m_{\eta
}^{2}}{Q^{2}+m_{\eta }^{2}}\right] +{\cal O}(Q^{4}),  \nonumber
\end{eqnarray}%
where the first term has been found in \cite{Venez79} and its chiral limit
follows the Crewther theorem \cite{Crew} maintaining that $\chi \left(
0\right) =0$ in any theory where at least one massless quark exists.

The estimates of $\chi ^{\prime }(0)$ existing in the literature are rather
controversial:
\begin{equation}
\chi ^{\prime }(0)=\left( 48\pm 6~{\rm MeV}\right) ^{2}\qquad \lbrack
5],\qquad \chi ^{\prime }(0)=\left( 26\pm 4~{\rm MeV}\right) ^{2}\qquad
\lbrack 6].  \label{Chi1iof}
\end{equation}%
Both estimates were found within the QCD sum rules method. These values of
the first moment of topological susceptibility have to be compared with the
value obtained in the Okubo-Zweig-Iizuka (OZI) case, the case free of axial
anomaly, which is
\[
\chi _{OZI}^{\prime }\left( 0\right) =\frac{f_{\pi }^{2}}{2N_{f}}\approx
\left( 39\quad {\rm MeV}\right) ^{2}.
\]

The principal point is that smallness of $\chi ^{\prime }(0)$\ is the base
for the one of the mechanisms explaining ''proton spin crisis'' problem \cite%
{NShVenez}. Indeed, within this approach it is assumed that the flavor
singlet axial charge $a_{0}\left( Q^{2}\right) $ is proportional to the
product of the first moment of the QCD topological susceptibility taken at
scale $Q^{2}$ and an RG-invariant coupling of `OZI Goldstone boson' with
nuclon
\begin{equation}
a_{0}(Q^{2})=\frac{1}{2m_{N}}6\sqrt{\chi ^{\prime }(0)}\widehat{\Gamma }%
_{\eta _{0}NN}.  \label{ShVa0}
\end{equation}%
This mechanism has been, however, criticized in \cite{EfrAnsL}. All this
makes important further model estimates of $\chi ^{\prime }(0)$.

Within the chiral quark model\footnote{%
The explicit calculations below are performed in $SU(2)$ sector of the model.%
} \cite{ADoLT00} based on the non-local structure of instanton QCD vacuum
\cite{DMikh} the full iso-singlet axial-vector vertex becomes \cite%
{Dorokhov:2003kf}
\begin{eqnarray}
\Gamma _{\mu 5}^{0}(k,q,k^{\prime }=k+q) &=&\left[ \gamma _{\mu
}-(k+k^{\prime })_{\mu }\frac{\left( \sqrt{M(k^{\prime })}-\sqrt{M(k)}%
\right) ^{2}}{k^{\prime 2}-k^{2}}\right.  \label{G50} \\
&&\left. -\frac{q_{\mu }}{q^{2}}2\sqrt{M(k^{\prime })M(k)}\frac{G^{\prime }}{%
G}\frac{1-GJ_{PP}(q^{2})}{1-G^{\prime }J_{PP}(q^{2})}\right] \gamma _{5}.
\nonumber
\end{eqnarray}%
where $M(k)$ is dynamical, momentum dependent quark mass, $G$ and $G^{\prime
}$ are 4-quark couplings in iso-triplet and iso-singlet channels,
correspondingly, and
\begin{equation}
J_{PP}(q^{2})\delta _{ab}=-\frac{i}{M_{q}^{2}}\int \frac{d^{4}k}{\left( 2\pi
\right) ^{4}}M\left( k\right) M\left( k+q\right) Tr\left[ S(k)\gamma
_{5}\tau ^{a}S\left( k+q\right) \gamma _{5}\tau ^{b}\right] .  \label{Ppp}
\end{equation}%
In (\ref{Ppp}) the (inverse) quark propagator is $S^{-1}(p)=\hat{p}-M(p)$.
Because of axial anomaly the singlet current does not contain massless pole,
since as $q^{2}\rightarrow 0$ one has:
\begin{equation}
\frac{1-GJ_{PP}(q^{2})}{-q^{2}}=G\frac{f_{\pi }^{2}}{M_{q}^{2}},
\label{NoGold}
\end{equation}%
where $f_{\pi }$ is pion weak decay constant and $M_{q}=M(0)$. The
cancellation of the massless pole occurs with help of the gap equation.
Instead, the current develops a pole at the $\eta ^{\prime }-$ meson mass
\footnote{%
See previous footnote.}, $1-G^{\prime }J_{PP}(q^{2}=-m_{\eta ^{\prime
}}^{2})=0$, thus solving the $U_{A}(1)$ problem. The vertex (\ref{G50})
satisfies the anomalous Ward-Takahashi identity:
\begin{equation}
q_{\mu }\Gamma _{\mu 5}^{(0)}(k,q,k^{\prime }=k+q)=\gamma
_{5}S_{F}^{-1}\left( k^{\prime }\right) +S_{F}^{-1}\left( k\right) \gamma
_{5}+\gamma _{5}\frac{2\sqrt{M(k^{\prime })M(k)}}{1-G^{\prime }J_{PP}(q^{2})}%
\left( 1-\frac{G^{\prime }}{G}\right) ,  \label{AnWTI}
\end{equation}%
where the last term is due to the anomaly. Thus, the QCD pseudoscalar
gluonium operator is interpolated by the pseudoscalar effective quark field
operator with coefficient expressed in terms of dynamical quark mass. This
is a consequence of the fact that in the effective quark model the
connection between quark and integrated gluon degrees of freedom is fixed by
the gap equation.

For completeness we display the vertex corresponding to the conserved
iso-triplet axial-vector current
\begin{eqnarray}
\Gamma _{\mu 5}^{a}(k,q,k^{\prime } &=&k+q)=T^{a}\left[ \gamma _{\mu
}-q_{\mu }\frac{M(k^{\prime })+M(k)}{q^{2}}-\right.  \label{GAtot} \\
&&\left. -(k+k^{\prime }-q\frac{k^{\prime 2}-k^{2}}{q^{2}})_{\mu }\frac{%
\left( \sqrt{M(k^{\prime })}-\sqrt{M(k)}\right) ^{2}}{k^{\prime 2}-k^{2}}%
\right] \gamma _{5}  \nonumber
\end{eqnarray}%
satisfying the axial Ward-Takahashi identity
\begin{equation}
q_{\mu }\Gamma _{\mu 5}^{a}(k,q,k^{\prime })=\gamma _{5}S_{F}^{-1}\left(
k^{\prime }\right) T^{a}+T^{a}S_{F}^{-1}\left( k\right) \gamma _{5}.
\label{AxWTI}
\end{equation}%
The axial-vector vertex (\ref{GAtot}) has a kinematical pole at $q^{2}=0$, a
property that follows from the spontaneous breaking of the chiral symmetry
in the limit of massless $u$ and $d$ quarks. Evidently, this pole
corresponds to the massless Goldstone pion.

The quark matrix elements of currents corresponding to vertices (\ref{G50})
and (\ref{GAtot}) can be expressed in terms of real form factors%
\begin{equation}
\left\langle p^{\prime }s^{\prime }\left\vert A_{\mu 5}^{(0,3)}\left(
0\right) \right\vert ps\right\rangle =\overline{u}_{s^{\prime }}\left(
p^{\prime }\right) T^{(0,3)}\left[ \gamma _{\mu }\gamma
_{5}G_{1}^{(0,3)}\left( q^{2}\right) -q_{\mu }\gamma _{5}G_{2}^{(0,3)}\left(
q^{2}\right) \right] u_{s}\left( p\right) ,  \label{Jme}
\end{equation}%
where $T^{(0,3)}=\left( 1,\tau ^{3}/2\right) $, $u_{s}\left( p\right) $ are
spinor solutions of the Dirac equation for free quarks, and the currents are
defined as%
\begin{equation}
A_{\mu 5}^{(0,3)}\left( q\right) =\int \frac{d^{4}k}{\left( 2\pi \right) ^{4}%
}\psi ^{\dagger }\left( k\right) \Gamma _{\mu 5}^{(0,3)}(k,q,k^{\prime
}=k+q)\psi \left( k+q\right) ,  \label{J}
\end{equation}%
with $\psi \left( k\right) $ being the solutions of the Dirac equation
\begin{equation}
\left[ \widehat{k}-M\left( k\right) \right] \psi \left( k\right) =0.
\label{DirEq}
\end{equation}%
By using the Dirac equation one gets%
\[
q^{\mu }A_{\mu 5}^{(3)}\left( q^{2}\right) =0,\qquad q^{\mu }A_{\mu
5}^{(0)}\left( q^{2}\right) =\frac{\left( 1-\frac{G^{\prime }}{G}\right) }{%
1-G^{\prime }J_{PP}(q^{2})}\int \frac{d^{4}k}{\left( 2\pi \right) ^{4}}2%
\sqrt{M(k^{\prime })M(k)}\psi ^{\dagger }\left( k\right) \gamma _{5}\psi
\left( k+q\right) .
\]%
Comparison with (\ref{Jme}) leads to the relations for form factors (taken
in the local limit $M(k)\approx M_{q}$)%
\[
G_{1}^{(3)}\left( q^{2}\right) =1,\qquad G_{2}^{(3)}\left( q^{2}\right)
=2M_{q}/q^{2},\qquad G_{1}^{(0)}\left( 0\right) =1,\qquad G_{2}^{(0)}\left(
0\right) =0
\]%
resembling the results for a model of free massive quarks.

Full model calculations lead to the following expression for the topological
susceptibility \cite{Dorokhov:2003kf}
\begin{eqnarray}
&&-\left( 2N_{f}\right) ^{2}\chi \left( Q^{2}\right)=  \nonumber \\
&=&2N_f\left( 1-\frac{ G^{\prime }}{G}\right) \left\{ Q^{2}J_{\pi A}\left(
Q^{2}\right) \left[ 1- \frac{G^{\prime }J_{AP}(Q^{2})}{M_{q}^{2}}+\frac{1}{%
1-G^{\prime }J_{PP}(Q^{2})}\right] \right.  \label{ChiMod} \\
&+&M_{q}^{2}J_{PP}\left( Q^{2}\right) \left( 1-\frac{G^{\prime }}{M_{q}^{2}}
J_{AP}(Q^{2})\right) \left[ \frac{GJ_{AP}(Q^{2})}{M_{q}^{2}}-\frac{
G-G^{\prime }}{G\left[ 1-G^{\prime }J_{PP}(Q^{2})\right] }\right]  \nonumber
\\
&+&\left. \frac{G}{M_{q}^{2}}\left[ 4N_{c}N_{f}\int \frac{d^{4}k}{\left(
2\pi \right) ^{4}}\frac{M\left( k\right) }{D\left( k\right) }\left[ M\left(
k\right) -\sqrt{M\left( k+Q\right) M\left( k\right) }\right] \right]
^{2}\right\} ,  \nonumber
\end{eqnarray}
where $D(k)=k^2+M^2(k)$ and the integrals $J_{AP}(q^2)$ and $J_{\pi A}(q^2)$
are defined by
\begin{eqnarray}
J_{AP}(q^{2}) &=&4N_{c}N_{f}\int \frac{d^{4}l}{\left( 2\pi \right) ^{4}}
\frac{M\left( l\right) }{D\left( l\right) }\sqrt{M\left( l+q\right) M\left(
l\right) },  \nonumber \\
J_{\pi A}\left( q^{2}\right) \delta _{ab}&=&\frac{q_{\mu }}{q^{2}}\int \frac{%
d^{4}k}{\left( 2\pi \right) ^{4}}Tr\left[ S(k)\widetilde{\Gamma }_{\mu
}^{5a}(k,q,k+q)S\left( k+q\right) \Gamma _{\pi }^{a}\left( k+q,k\right) %
\right] .  \nonumber
\end{eqnarray}
At large $Q^{2}$ one obtains the power-like behavior consistent with the OPE
prediction (\ref{ChiQCDlarge}), namely
\begin{equation}
-\left( 2N_{f}\right) ^{2}\chi \left( Q^{2}\rightarrow \infty \right) =\frac{
2N_fM_{q}^{2}}{G}\left( 1-\frac{G^{\prime }}{G}\right) .  \label{Chi1As}
\end{equation}

At zero momentum the topological susceptibility is zero
\begin{equation}
\chi \left( 0\right) =0,  \label{CrewTh}
\end{equation}%
in accordance with the Crewther theorem. For the first moment of the
topological susceptibility we obtain \cite{Dorokhov:2003kf}
\begin{equation}
\chi ^{\prime }\left( 0\right) =\frac{1}{2N_{f}}\left\{ f_{\pi }^{2}\left( 2-%
\frac{G^{\prime }}{G}\right) +\left( 1-\frac{G^{\prime }}{G}\right)
^{2}J_{AP}^{\prime }(0)\right\} .  \label{Chi1Mod}
\end{equation}

If the OZI rule were exact in the flavor singlet channel and there were no
anomaly, then one would have $G^{\prime }=G$ and $\chi ^{\prime }\left(
0\right) =\chi _{OZI}^{\prime }\left( 0\right) $. But actually one has
strong attraction in the iso-triplet channel and strong repulsion due to the
anomaly in the iso-singlet channel that means that one has always inequality
$G^{\prime }<G.$ The second, negative term in (\ref{Chi1Mod}) is numerically
suppressed with respect to the first, positive term, $J_{AP}^{\prime
}(0)/f_{\pi }^{2}\approx -0.24$. Thus, from the existence of the anomaly we
always have inequality%
\begin{equation}
\chi ^{\prime }\left( 0\right) >\chi _{OZI}^{\prime }\left( 0\right) ,
\label{ChiIneq}
\end{equation}%
and it is impossible to get anomalously small $\chi ^{\prime }\left(
0\right) $. At this point we also mention other alternative approaches to
the spin crisis problem based on screening of topological charge in the QCD
vacuum \cite{forte,DKspin90} (see for review, {\it e.g.} \cite{DKspinRev}).

The constants $G$ and $G^{\prime }$ are fixed with the help of the meson
spectrum. Approximately one has $G^{\prime }\approx 0.1~G$. As a profile for
the dynamical quark mass we take a Gaussian form
\begin{equation}
M(u)=M_{q}\exp \left( -2u/\Lambda ^{2}\right)
\end{equation}
with the model parameters $M_{q}=0.3~{\rm GeV,}\qquad \Lambda =1.085~{\rm GeV%
}. $ Then the estimate for the first moment of the topological
susceptibility is \cite{Dorokhov:2003kf}
\begin{equation}
\chi ^{\prime }(0)=\left( 50~{\rm MeV}\right) ^{2}.  \label{Chi1Est}
\end{equation}%
To get the above result we have taken $N_{f}=3$ in Eq. (\ref{Chi1Mod}). We
can see that the model gives the value of $\chi ^{\prime }(0)$\ which is
close to the estimate of Ref.~\cite{IoOg98}. The influence of the current
quark masses on $\chi ^{\prime }(0)$ is expected to be small and the
contribution of $\pi $- and $\eta $-mesons may be found from Eq. (\ref%
{ChiQCDsmall})
\[
\chi _{\pi ,\eta }^{\prime }(0)\approx \left( 28~{\rm MeV}\right) ^{2},
\]
and thus $\chi _{tot }^{\prime }(0)\approx \left( 57~{\rm MeV}\right) ^{2}$
for the total result.

The model prediction for the topological susceptibility is shown in Fig. 1.
In the region of small and intermediate momenta our result is quantitatively
close to the prediction of the QCD sum rules with the instanton effects
included \cite{IofSams00}.
%%%%%%%%%%%%%%%%%%%%%%%%%%%%%%%%%%%%%%%%%%%%%%%%%%

\begin{wrapfigure}{R}{8cm}
\mbox{\epsfig{figure=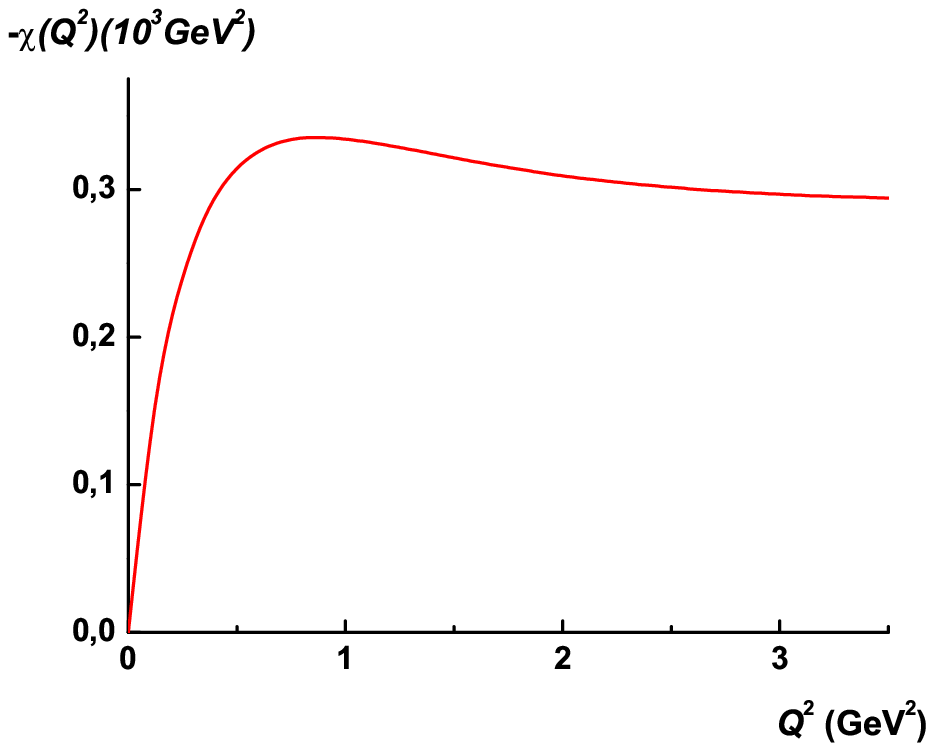,width=7.8cm,height=6cm}} {\small{\bf Figure 1.} Topological susceptibility versus
$Q^2$ predicted by the model with $G^{\prime}=0.1~G$, Eq. (\protect\ref{ChiMod}).}
%\medskip
\label{f3}
\end{wrapfigure}
%%%%%%%%%%%%%%%%%%%%%%%%%%%%%

%\begin{figure}[]
%\includegraphics[width=10cm,height=7cm]{FigTopB.eps}
%\caption{Topological susceptibility versus $Q^2$ predicted by the model with
%$G^{\prime}=0.1~G$, Eq. (\protect\ref{ChiMod}), (solid line).}
%\end{figure}

In the present talk we analyzed the correlation function of the singlet
axial-vector currents within an effective non-local chiral quark model. By
considering this correlator the topological susceptibility was found as a
function of the Euclidean-momentum and its first moment was estimated. We
demonstrated that in realistic situation one always gets the inequality $%
\chi ^{\prime }(0)>\chi _{OZI}^{\prime }(0)$, thus discarding the mechanism
explaining the 'spin crisis' based on anomalous smallness of $\chi ^{\prime
}(0).$ In addition, the fulfillment of the Crewther theorem was
demonstrated. It would be interesting to verify the predictions given in
Fig. 1\ref{f3} by modern lattice simulations.

The author is grateful to W. Broniowski for fruitful cooperation and S.B.
Gerasimov, H. Forkel, N. I. Kochelev, S.V. Mikhailov and O.V. Teryaev for
useful discussions on the subject of the present work. This work is
partially supported by grants RFBR (nos. 02-02-16194, 03-02-17291) and
INTAS-00-00-366.

%\newpage

\end{document}